\documentclass[12pt,preprint]{aastex}

\shorttitle{Shallow Decay of GRBs} \shortauthors{Mao \& Wang}

\begin{document}

\title{A New Scenario on X-ray Shallow Decay of Gamma-ray Bursts}

\author{Jirong Mao and Jiancheng Wang}
\affil{Yunnan Observatory, National Astronomical Observatories,
Chinese Academy of Sciences, P.O. Box 110, Kunming, Yunnan
Province, 650011, China}

\email{jirongmao@ynao.ac.cn}

\begin{abstract}
In this Letter, we propose that a microphysical process takes a
vital role in the shocked region in which the prompt emission of
GRBs is emitted. The turbulent energy is included in the internal
energy transferred by the kinetic energy of the shock. It
dissipates through stochastic acceleration for the electrons to
supply the early X-ray emission in the phase of shallow decay. We
put the constraints on the time evolution of microphysical
parameters. The early X-ray fluxes can be obtained by this
scenario and these results are consistent with the Swift
observation.
\end{abstract}


\section{Introduction}
In the Swift era \citep{gehrels04}, the X-ray telescope (XRT)
observation \citep{burrows05} provides the complete light curves
of gamma-ray bursts (GRBs) in the 0.2-10 keV band. One of the most
interesting discoveries is the so-called shallow decay segment:
the flux plateau of $F\propto t^{-0.5}$ within $10^3-10^4$ second
after the trigger \citep{campana05,depasquale06}. Recent statistic
analyses \citep{nousek06,obrien06,liang07} have revealed that the
phase of shallow decay in the X-ray afterglow might be a common
feature of the long GRBs. The shallow decay in the early X-ray
light curve is still a mystery, although theoretical explanations
have been put forward from several aspects (see Zhang 2007 for a
comprehensive review). Most of the models are: hydrodynamics of
the shock by energy injection \citep{granot06a,zhang06}, geometry
of the jet \citep{eichler06,toma06}, varying microphysical
parameters \citep{fan06,granot06b,panaitescu06,ioka06}, late
prompt emission \citep{ghisellini07} or up-scattered forward-shock
emission \citep{panaitescu07}.

From the point of microphysics in the hydrodynamic evolution, the
nature of coupling between the electrons, protons and magnetic
field is complex \citep{chiang99}. Usually the simple way is to
assume the equipartition between electrons, protons and magnetic
field \citep{panaitescu98}. However, in this Letter, we propose
that the kinetic energy of the relativistic shocks in the plasma
has been converted to the internal energy as three parts: (1)
$\varepsilon_B$, which is the energy of magnetic field; (2)
$\varepsilon_e$ and $\varepsilon_p$, it means that the electrons
and protons/positrons are accelerated by the shocks, normally,
this is the process of first-order Fermi acceleration; (3)
$\varepsilon_t$, which presents the turbulent energy, and this
energy would sustain a relatively long time (see Section 2). The
last part has not been taken into account by the former research.
In our novel scenario, the relativistic electrons accelerated by
the first-order Fermi acceleration emit the gamma-ray by
synchrotron radiation, after the decrease of the prompt emission
tail which is shown as the deep decay in the early X-ray light
curve, indicating that the internal shocks are abated, the
follow-up turbulence and its effects might be dominated. The
turbulence could transfer its energy to the electrons via
second-order Fermi acceleration, which is also called as the
stochastic acceleration. Therefore, in this turbulent region, due
to the resonant interaction between electrons and plasma waves,
the electrons buried in the magnetic field are re-accelerated by
the stochastic acceleration. The emission of these electrons
dominates the shallow decay phase, until the turbulent energy
dissipates and the external shock sweeps the surround medium thus
the deep decay appears again.

In Section 2, we review the Fokker-Planck equation and list the
coefficients associated with the turbulent term. In Section 3, the
turbulent parameter $\varepsilon_t$, as same as $\varepsilon_e$
and $\varepsilon_B$, is introduced. Due to the turbulence, these
microphysical parameters evolved with time are constrained by the
process of stochastic acceleration. Finally, we select these
relations to reproduce the feature of shallow decay. The
discussions are given in Section 4.

\section{Stochastic Acceleration}
The stochastic acceleration was suggested as an non-neglected
mechanism to produce the high energy particles in GRBs
\citep{waxman95,dermer01}. The numerical simulation has confirmed
that the stochastic acceleration in the relativistic shocks plays
an important role on the particle energy distribution and
evolution \citep{virtanen05}.

In general, the charge particles are expected to be accelerated
through resonant interactions with the magnetized plasma waves.
For this stochastic acceleration, particle distribution function
$f(\gamma,t)=dN(\gamma,t)/d\gamma$ satisfies the Fokker-Planck
equation, which is the kinetic equation of single electron in the
energy space:
\begin{equation}
\frac{\partial f(\gamma,t)}{\partial t}=\frac{\partial^2}{\partial
\gamma^2}[D(\gamma)f(\gamma,t)]-\frac{\partial}{\partial
\gamma}[A(\gamma)+ \dot{E}_L f(\gamma,t)]-
\frac{f(\gamma,t)}{T(\gamma)}+Q(\gamma,t)
\end{equation}
where $A(\gamma)$ is the acceleration rate, $D(\gamma)$ the
diffusion rate, $\dot{E_L}$ energy loss rate, $T(\gamma)$ the mean
escape time of electrons and $Q(\gamma,t)$ the source term. The
turbulent spectrum is usually employed by the form of Kolmogorov
or Kraichnan as $W(k)\propto k^{-q}$ \citep{zhou90}.

\citet{park95} already explored the various analytic solutions and
they illustrated that the coefficients of the equation are related
to the microphysics of turbulent plasma. In that paper, they
offered one special example, the so-called hard-sphere
approximation, where the index of turbulent spectrum $q=2$. This
detailed work has been complemented by \citet{becker06}, in which
the time-dependent Green's function was reexamined and the case of
$q<2$ was considered. However, the energy loss, such as
synchrotron or inverse Compton, was not included in the solution.

\citet{wang01} calculated the equation which is applied in the
spectral variability of blazars. Recently, \citet{liu06a} studied
the stochastic acceleration in Sagittarius A* adopting $q=2$.
Under the physical conditions of Sagittarius A*, the steady state
of $f(\gamma)$ was derived \citep{liu06b}. \citet{manolakou07}
have obtained the time dependent solution of $f(\gamma)$ taking
into account the cooling via synchrotron and inverse Compton
radiation. In this Letter, the calculation is simply on the
coefficients and we avoid to give any details about $f(\gamma)$.

In order to explain the shallow decay phase of GRB X-ray
afterglow, assuming (i) the scatter path is much less then the
length of the turbulent region, (ii) there is no continual
injection and the accelerated electrons have the synchrotron
cooling thus the steady state exists, we have the simplified
Fokker-Planck equation of $\partial f/\partial t=0$ and all the
coefficients are time-independent. Here, we repeat the formulae
produced by \citet{dermer96} and \citet{wang02}:
\begin{equation}
\dot{E}_L=\frac{4\sigma_T}{3m_ec}(U_B+U_{ph})(\gamma\beta)^2,
\end{equation}
\begin{equation}
A(\gamma)=\frac{\pi}{2}\frac{q-1}{q}\frac{U_t}{U_B}\beta_g^2ck(r_Lk)^{q-2}p^{q-1},
\end{equation}
\begin{equation}
D(\gamma)=\frac{\pi}{2}\frac{q-1}{q(q+2)}\frac{U_t}{U_B}\beta_g^2ck(r_Lk)^{q-2}p^q\beta^{-1},
\end{equation}
\begin{equation}
T(\gamma)=\frac{\pi}{2}\frac{q-1}{q}\frac{U_t}{U_B}ck(r_Lk)^{q-2}p^{q-2}(\Delta
t)^2\beta^{-1},
\end{equation}
where $\beta=v/c$, $\gamma$ is the Lorentz factor of electron,
$k=(ct)^{-1}$ the wavenumber, $r_L=m_ec^2/eB$ the nonrelativistic
Larmor radius of the electron, $\sigma_T$ the Thomson cross
section, \textbf{p} dimensionless momentum. $U_B$, $U_t$ and
$U_{ph}$ are the magnetic energy density, turbulent energy
density, and photon field energy density respectively. Suppose the
the particles are accelerated by Alfv\'{e}n turbulence, the group
speed $\beta_g$ is equal to Alfv\'{e}n speed.

In the traditional models, two microphysical parameters, the ratio
of magnetic field energy to the total internal energy
$\varepsilon_B=U_B/U$ and the fraction of the total internal
energy that goes into the random motions of electrons
$\varepsilon_e=U_e/U$, are drawn into the study of GRBs
\citep{sari96}. In our scenario, the turbulent energy is included
in the total internal energy. Similar to the $\varepsilon_e$ and
$\varepsilon_B$, we define the dimensionless parameter
$\varepsilon_t$ to present the turbulent property as:
\begin{equation}
\varepsilon_t=\frac{U_t}{U}.
\end{equation}

Since the dominant emission is synchrotron radiation during the
shallow decay phase, the term of photon field $U_{ph}$ which is
associated with the inverse Compton process can be ignored. Of the
single electron, the stochastic acceleration rate is balanced to
the energy loss rate through synchrotron radiation:
\begin{equation}
A(\gamma)=\dot{E}_L.
\end{equation}

The energy ratio of accelerated electrons in the turbulent region
is equal to the total stochastic acceleration rate:
\begin{equation}
\frac{d\varepsilon_e}{dt}= \frac{\int
A(\gamma)f(\gamma)d\gamma}{\int f(\gamma)d\gamma}.
\end{equation}

The dissipation timescale of turbulent energy can be estimated by:
\begin{equation}
T_d=\frac{\varepsilon_t U}{<\dot{E}_L>}=\frac{\alpha
U_B}{d\varepsilon_e/dt},
\end{equation}
where $\alpha=\varepsilon_t/\varepsilon_B$. The relativistic
electrons re-accelerated by stochastic acceleration are assumed to
have energy distribution $f(\gamma)\propto \gamma^{-p}$ with the
limits of $[\gamma_{min}, \gamma_{max}]$ and produce X-ray
emission. For $p=2.2$, the timescale $T_d$ is given by:
\begin{equation}
T_d=8.6\times10^3\times\alpha\times
(\frac{\Gamma}{100})^{-1.6}(\frac{\varepsilon_B}{0.001})^{0.2}(\frac{n}{10^2})^{0.2}
(\frac{\gamma_{min}}{10.0})^{-1.2}~s,
\end{equation}
where $\Gamma$ is the bulk lorentz factor, $\gamma_{min}$ is the
minimum lorentz factor of the accelerated electron by turbulence
and $n$ is the number density of shocked medium. The dissipation
timescale of the turbulent energy is coincident with the
observational timescale of shallow decay phase.

Therefore, in the shocked region with turbulence, the evolution of
microphysical parameters $\varepsilon_e$, $\varepsilon_B$ and
$\varepsilon_t$ can be derived from the coefficients of
Fokker-Planck equation. From equation (7) and (8), we obtain:
\begin{equation}
\varepsilon_B\propto \varepsilon_t^{2/q}t^{2(1-q)/q}
\end{equation}
and
\begin{equation}
\frac{d\varepsilon_e}{dt}\propto
\varepsilon_t\varepsilon_B^{(2-q)/2}t^{1-q}.
\end{equation}

Furthermore, in this turbulent region, we assume that these
microphysical parameters have the forms of $\varepsilon_e\propto
t^a$, $\varepsilon_B\propto t^b$ and $\varepsilon_t\propto t^d$.
We insert them into equation (11) and (12) and put the constraints
on a, b and d by two algebra equations:
\begin{equation}
b=2d/q+2(1-q)/q
\end{equation}
and
\begin{equation}
a-1=b(2-q)/2+(1-q)+d.
\end{equation}
We choose the turbulent spectrum of Kraichnan with the index
$q=4/3$. All of the possible values of $a$, $b$ and $d$ are shown
in Figure 1, 2 and 3.

From the standard afterglow model, the fluxes in the early X-ray
band were written by \citet{sari98}. There are two different
limits: adiabatic and radiative case. For explanation of the
emission in shallow decay phase, we obtain the early X-ray light
curve under the adiabatic case as:
\begin{equation}
F\propto\varepsilon_e^{p-1}\varepsilon_B^{(p-2)/4}t^{(2-3p)/4}\propto
t^{(bp-2b-3p+2)/4+a(p-1)}.
\end{equation}
As an example, we adopt the index $q=4/3$ and $p=2.2$. Four
evolutionary fluxes are achieved in the table~\ref{tbl-1}, they
are corresponding to the minimum and maximum values of $a$ and
$b$. While for the radiative case, the X-ray flux is:
\begin{equation}
F\propto\varepsilon_e^{p-1}\varepsilon_B^{(p-2)/4}t^{(2-6p)/7}\propto
t^{b(p-2)/4+(2-6p)/7+a(p-1)}.
\end{equation}
Table~\ref{tbl-2} lists the possible fluxes evolved with the time.

\section{Discussions}
Due to the turbulent energy and its dissipation in the shocked
region, after the gamma-ray radiation, the electrons are
re-accelerated by the stochastic acceleration and emit the early
X-ray fluxes. Therefore, the energy injection in the central
engine is not required to reproduce the emission in the shallow
decay phase. Our results manifest that the temporal feature in the
shallow decay phase represents the process of turbulent
dissipation. The microphysical parameters, $\varepsilon_e$,
$\varepsilon_B$ and $\varepsilon_t$, are varied with the time in
the shocked region. Given the special values of $p=2.0$ and
$a=0.5$ in the adiabatic case, our calculation can represent the
typical observational flux as $F\propto t^{-0.5}$. Moreover, it is
noted that the same shallow decay phase is also detected in the
optical band as well \citep{mason06}, our simple interpretation is
that the synchrotron emission of X-ray band and optical band may
be original from the same shocked region with turbulent energy.
Another advantage of our model is that the crisis of radiative
efficiency \citep{lloyd04,ioka06,fan06,zhang07} is therefore
dispelled without any additional assumption of ejection/ejecta
from the central engine. From this point of view, we support the
internal shock pattern of standard fireball model.

In this work, we assume $f(\gamma)\propto \gamma^{-p}$ where
$p=2.2$ as the typical value. The detailed calculation of the
energy distribution $f(\gamma)$ is not needed, because in our
scenario the timescale of acceleration is much smaller than that
of turbulent energy dissipation. Since the timescale of
acceleration is also smaller than that of shock hydrodynamics,
therefore, during the phase of shallow decay, the index of
spectrum does not change \citep{liang07}.

Generally, the varied values of $\varepsilon_e$ and
$\varepsilon_B$ lead to the different hydrodynamic evolution and
emissions in the entire afterglow \citep{mao01a,mao01b}. In this
Letter, the early X-ray fluxes in the shallow decay phase are
estimated either in the adiabatic or radiative case through the
turbulent process. The average slope of radiative case is deeper
than that of adiabatic case. Compared with the XRT observation,
the hydrodynamic evolution of adiabatic case could be the realized
regime during the shallow decay phase. Although it is hard to
obtain the values of $\varepsilon_e$ and $\varepsilon_B$ in our
scenario, we put the constraints of $\varepsilon_e\ll 1$ and
$\varepsilon_B\ll 1$. More observational samples are particularly
requested for the further investigations.

\acknowledgments We thank the staff of Swift group in Astronomico
di Brera (Merate), G. Ghisellini and Z. G. Dai for the general
discussion. This work is financially supported by the Chinese
National Science Fund 10673028.

\clearpage

\begin{figure}
\epsscale{.80} \plotone{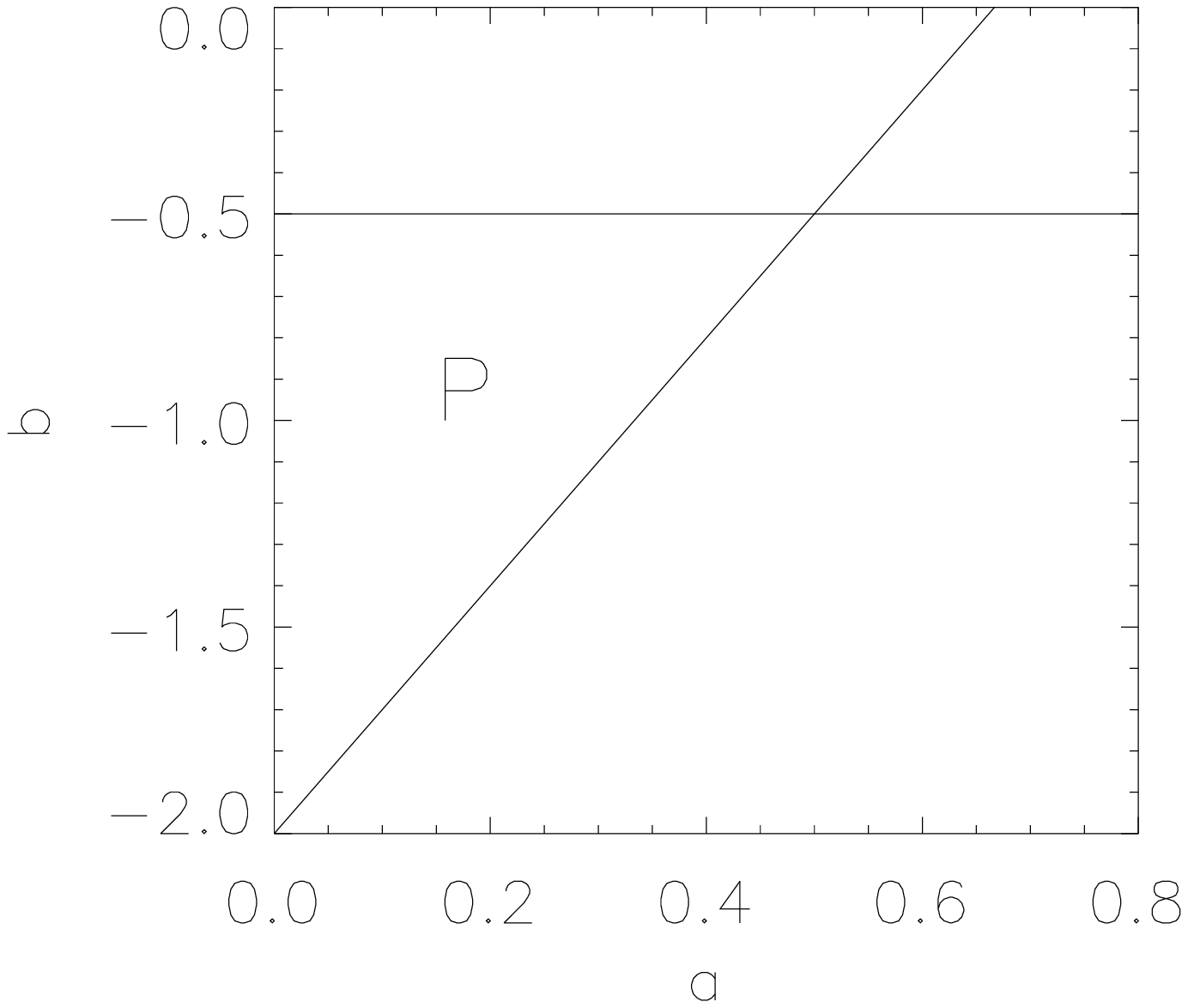} \caption{The a-b diagram. The
related values of a and b are shown in the region noted by the
capital \textsc{P}. \label{fig1}}
\end{figure}

\clearpage

\begin{figure}
\epsscale{.80} \plotone{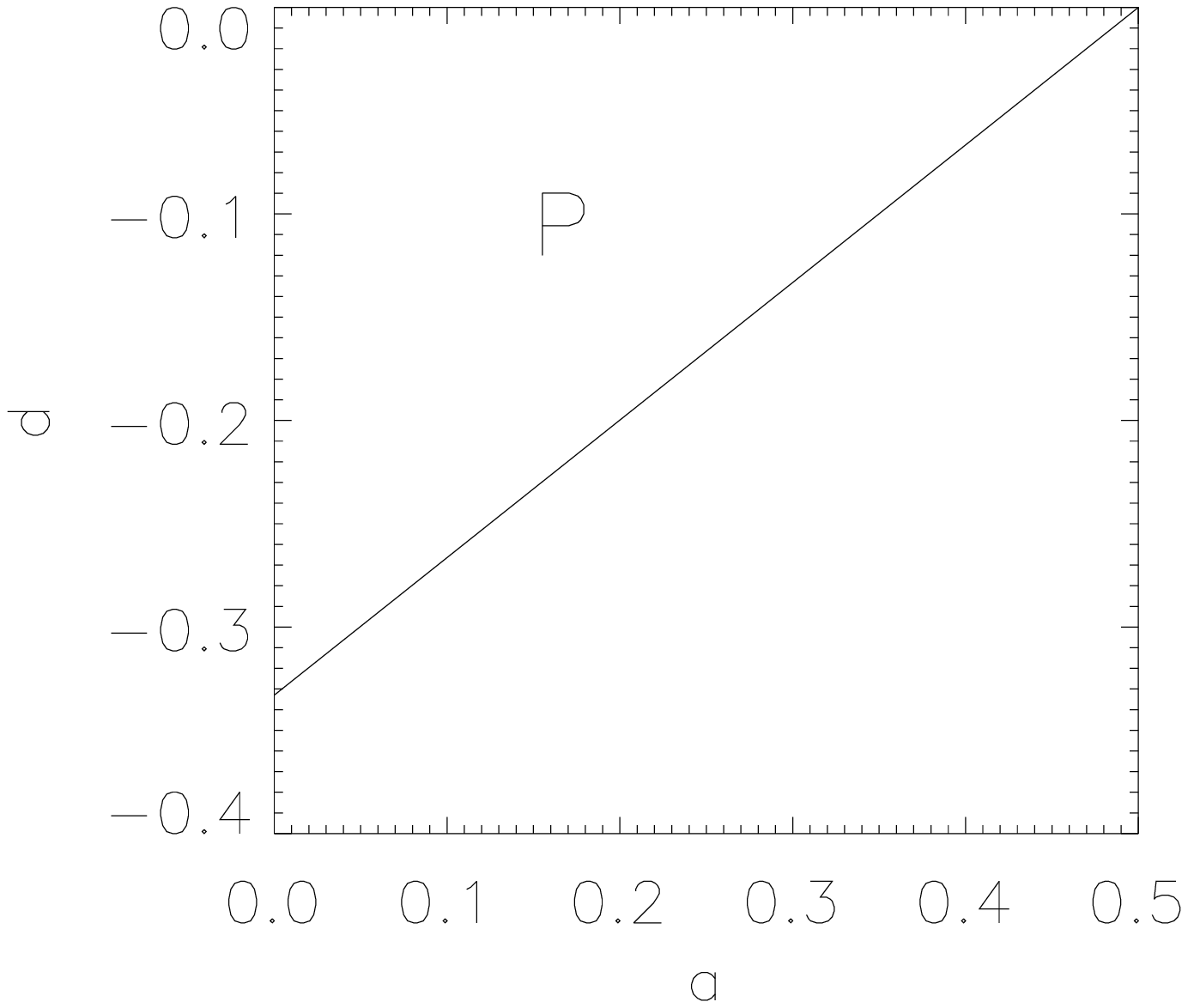} \caption{The a-d diagram. The
related values of a and d are shown in the region noted by the
capital \textsc{P}. \label{fig2}}
\end{figure}

\clearpage

\begin{figure}
\epsscale{.80} \plotone{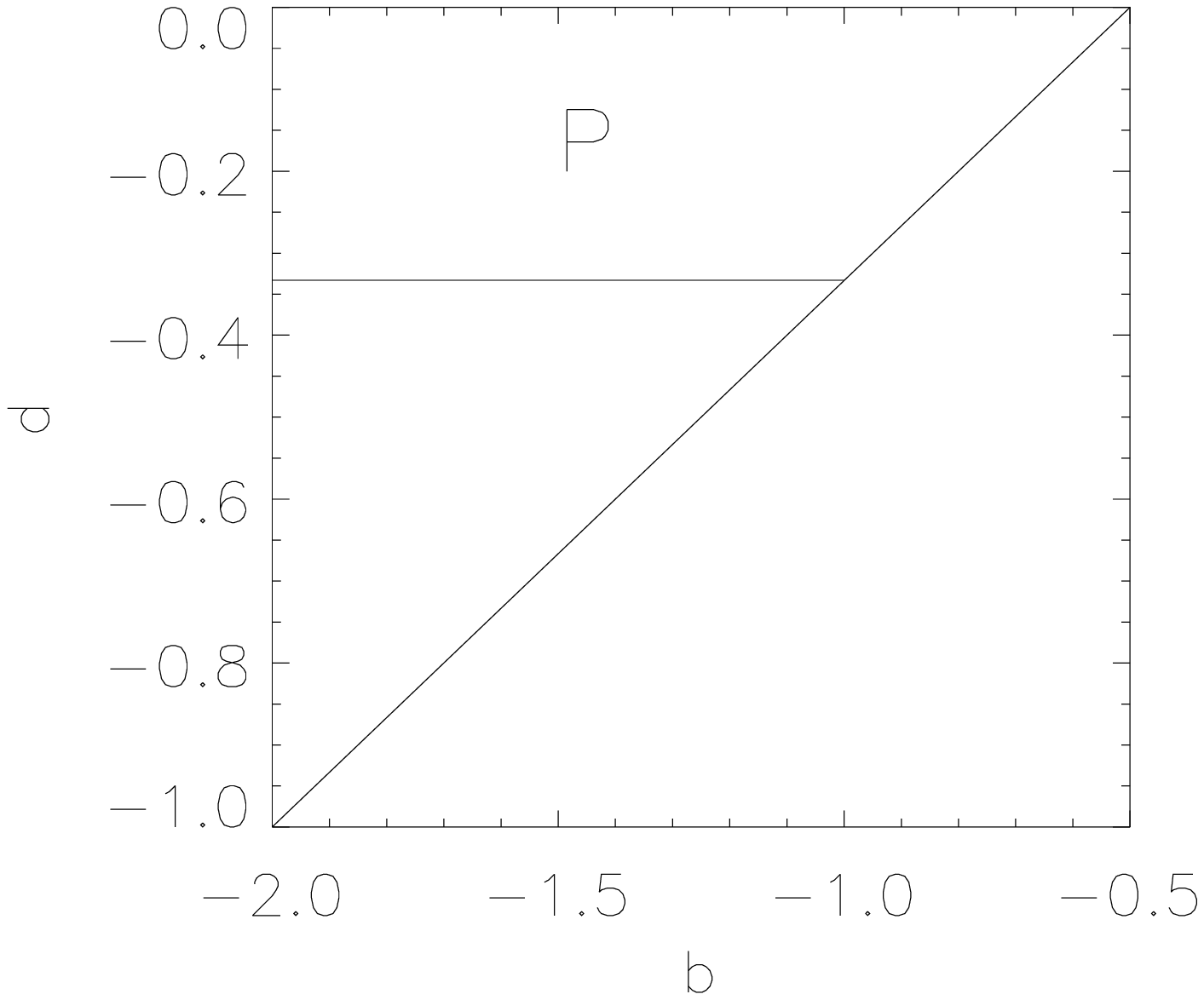} \caption{The b-d diagram. The
related values of b and d are shown in the region noted by the
capital \textsc{P}. \label{fig3}}
\end{figure}

\clearpage

\begin{table}
\begin{center}
\caption{The limit fluxes of shallow decay for the adiabatic
case\label{tbl-1}}

~

\begin{tabular}{crrrr}
\tableline\tableline adiabatic case& $a_{min}=0$ & $a_{max}=0.5$ & \\
\tableline
$b_{min}=-2$   & $F\propto t^{-1.25}$  & $F\propto t^{-0.65}$ &\\
$b_{max}=-0.5$ &~~~~~~$F\propto t^{-1.175}$ &~~~~~~~$F\propto t^{-0.575}$ & \\
\tableline
\end{tabular}

~

~

~

\caption{The limit fluxes of shallow decay for the radiative
case\label{tbl-2}}

~

\begin{tabular}{crrrr}
\tableline\tableline radiative case& $a_{min}=0$ & $a_{max}=0.5$ & \\
\tableline
$b_{min}=-2$   & $F\propto t^{-1.5}$  & $F\propto t^{-0.9}$ &\\
$b_{max}=-0.5$ &~~~~~~$F\propto t^{-1.575}$ &~~~~~~~$F\propto t^{-0.975}$ & \\
\tableline
\end{tabular}
\end{center}

\end{table}

\end{document}